\documentclass[pra,floatfix,showpacs,twocolumn,nofootinbib]{revtex4-1}

\usepackage{graphicx,graphics,epsfig,times,bm,bbm,amssymb,amsmath,amsfonts,mathrsfs,color}
\usepackage{hyperref}
\usepackage[normalem]{ulem}

\newcommand{\beq} {\begin{equation}}
\newcommand{\eeq} {\end{equation}}
\newcommand{\bea} {\begin{eqnarray}}
\newcommand{\eea} {\end{eqnarray}}
\newcommand{\bes} {\begin{subequations}}
\newcommand{\ees} {\end{subequations}}

\newcommand{\ignore}[1]{}
\def\ket#1{{|#1\rangle}}
\def\bra#1{{\langle#1|}}
\newcommand{\mc}[1]{\mathcal{#1}}
\newcommand{\vac}{\ket{\mathbf{0}}}
\newcommand{\vacb}{\bra{\mathbf{0}}}

\begin{document}

\title{No-go theorem for passive single-rail linear optical quantum computing}

\author{Lian-Ao Wu}
\affiliation{Ikerbasque--Basque Foundation for Science and Department of Theoretical
Physics and History of Science, The Basque Country University (EHU/UPV), PO Box 644, 48080 Bilbao, Spain}
\author{Philip Walther}
\affiliation{Faculty of Physics, University of Vienna, Boltzmanngasse 5, Vienna A-1090, Austria}
\author{Daniel Lidar}
\affiliation{Departments of Electrical Engineering, Chemistry, and Physics,
Center for Quantum Information \& Technology,
University of Southern California, Los Angeles, California 90089, USA}

\begin{abstract}
Photonic quantum systems are among the most promising architectures for quantum computers.
It is well known that for dual-rail photons effective non-linearities and near-deterministic non-trivial two-qubit gates can be achieved via the measurement process and by introducing ancillary photons. While in principle this opens a legitimate path to scalable linear optical quantum computing, the technical requirements are still very challenging and thus  other optical encodings are being actively investigated. One of the alternatives is to use single-rail encoded photons, where entangled states can be deterministically generated. Here we prove that even for such systems universal optical quantum computing using only passive optical elements such as beam splitters and phase shifters is not possible. This no-go theorem proves that photon bunching cannot be passively suppressed even when extra ancilla modes and arbitrary number of photons are used. Our result provides useful guidance for the design of optical quantum computers.
\end{abstract}

\maketitle

\section{Introduction}
Optical implementations of qubits have played an important role in quantum information science~\cite{OBrien:07,Aspuru:12}. Photons exhibit an intrinsic lack of decoherence and are simple to control by standard off-the-shelf components. Furthermore, photonic qubits for quantum computation are particularly attractive because they can be used to interface to various
quantum communication applications\cite{OBrien:09}. Due to the extremely small photon-photon coupling available in existing materials, it was at one point believed that optical qubits could {not} be used for scalable quantum computation. However, it is now understood that the process of photon detection itself can lead to effective photon-photon nonlinearities. In particular Knill, Laflamme, and Milburn (KLM)\cite{Knill:00} {launched the field of linear optics quantum computing (LOQC) by
showing} that deterministic single-photon sources and
high-efficiency single-photon detectors allow the realization of scalable{, probabilistic}
quantum computation purely with linear optical elements. {This holds in spite of the fact that using linear optics alone, the success probability of the nonlinear sign shift gate used in the KLM scheme cannot be improved above $1/4$\cite{Scheel:04,Eisert:05}.} {Since the original KLM
proposal, a number of authors have suggested various simplifications,
modifications, and {optimizations}\cite{Pittman:01,Knill:02,Ralph:01a,Uskov:09}. Although these
results formally show that scalable quantum computing is
possible, the realization is very demanding
in practice due to the large resource overhead arising from the
required non-deterministic
{photon detection events}. This becomes
particularly apparent when considering beamsplitter-based
two-photon gates, which are probabilistic in nature. In fact all optical
two-qubit gates\cite{Kok:07}, including the promising non-destructive CNOT
gate\cite{Pittman:02,Pittman:03}, fail in the case where more than one
photon is emitted into the same optical output mode, due
to photon bunching. One proposal
to actively suppress
such gate failure events in the polarization encoding uses
the quantum Zeno effect, by coupling to the output
light fields to atomic transitions\cite{Franson:04,Leung:07}.

{While the previous discussion concerned mostly dual-rail qubits, less is known about single-rail photonic qubits.} Here we address the natural question of whether
linear optical elements alone can be used to establish quantum interference
such that the photon-bunching effect, and thus the
probability of
gate failure, can be {eliminated for the case of single-rail encoding}.
We show that this cannot be achieved
in the sense that one cannot simultaneously {and deterministically} implement a linear optical two-photon entangling gate and decouple the double-occupancy states (in this work we are not concerned with single-photon nonlocality\cite{PhysRevLett.99.180404}). We thus prove a no-go theorem for photon bunching suppression via all-unitary linear optics. Our result is complementary to the recent proof that the
  single-photon fraction in any of the single-mode states resulting from purely
  linear optical processing (even including conditioning on results of
  detections) cannot be made to exceed the efficiency of the best available
  photon source\cite{Berry:10}. We stress that our proof applies in the setting of deterministic LOQC; we do not address the possibility of simultaneously optimizing the suppression of photon bunching and the success probability of LOQC gates in the single-rail setting. Nor do we address here what may be gained by adding measurements or other non-unitary operations\cite{Verstraete:2009fk}.

\section{Results}
\subsection{Problem formulation}
In the standard circuit model of quantum computing any unitary transformation on $n$ qubits can be
decomposed as a product of gates, each of which acts nontrivially on at most two qubits, and is
the identity on the other qubits\cite{Nielsen:book}. Likewise, in the linear optics model, any unitary transformation
on $M$ modes can be decomposed into a product of ``linear optical elements,'' each of which acts nontrivially
on at most two modes, and is the identity on the other $M-2$ modes\cite{Reck:94,Aharonson:10}.
As linear optical operations we consider just \emph{passive} elements, in particular
phase shifters and beam splitters,
as they generate all linear optical elements [i.e., all $2 \times 2$ unitaries, the group U$(2)$] when acting on the same pair of modes.
Moreover, when they are allowed to act on overlapping pairs of modes, the linear optical
elements generate all $M \times M$ unitaries [i.e., the group U$(M)$]\cite{Reck:94}.

Single photon optical qubits come primarily in two varieties: single-rail\cite{Pegg:98,Lee:00}, where a qubit is represented by the absence or presence of a single photon of fixed polarization in one optical mode, and dual-rail\cite{Nielsen:book}, where each qubit is encoded into the presence of a single photon in one or the other of two spatial optical modes. Polarization qubits\cite{Milburn:89}---where each qubit is represented by two orthogonal polarization modes---are formally equivalent to dual-rail qubits\cite{Kok:07}, so we will use both interchangeably (see Figure~1). Most LOQC proposals, including the original KLM scheme, use dual-rail qubits, but there is considerable interest in the single-rail encoding scheme as well, as evidenced by the numerous experiments devoted to
preparation of arbitrary {states of}
single-rail qubits\cite{Lombardi:02,Resch:02,Lvovsky:02,Babichev:03,Lvovsky:04}.  In the dual-rail encoding single-qubit operations are straightforward while non-trivial two-qubits are challenging and are implemented probabilistically\cite{Knill:00}. The reverse is true for single-rail encoding:
{it is easy to generate two-qubit entanglement deterministically,} while single-qubit gates can be implemented probabilistically\cite{Lund:02}. In both the single and dual-rail encodings two-qubit gate failure due to photon bunching is a major challenge standing in the way of scalability. {Let us next explain this in more detail.}

\begin{figure*}
\includegraphics[width=0.75\textwidth]{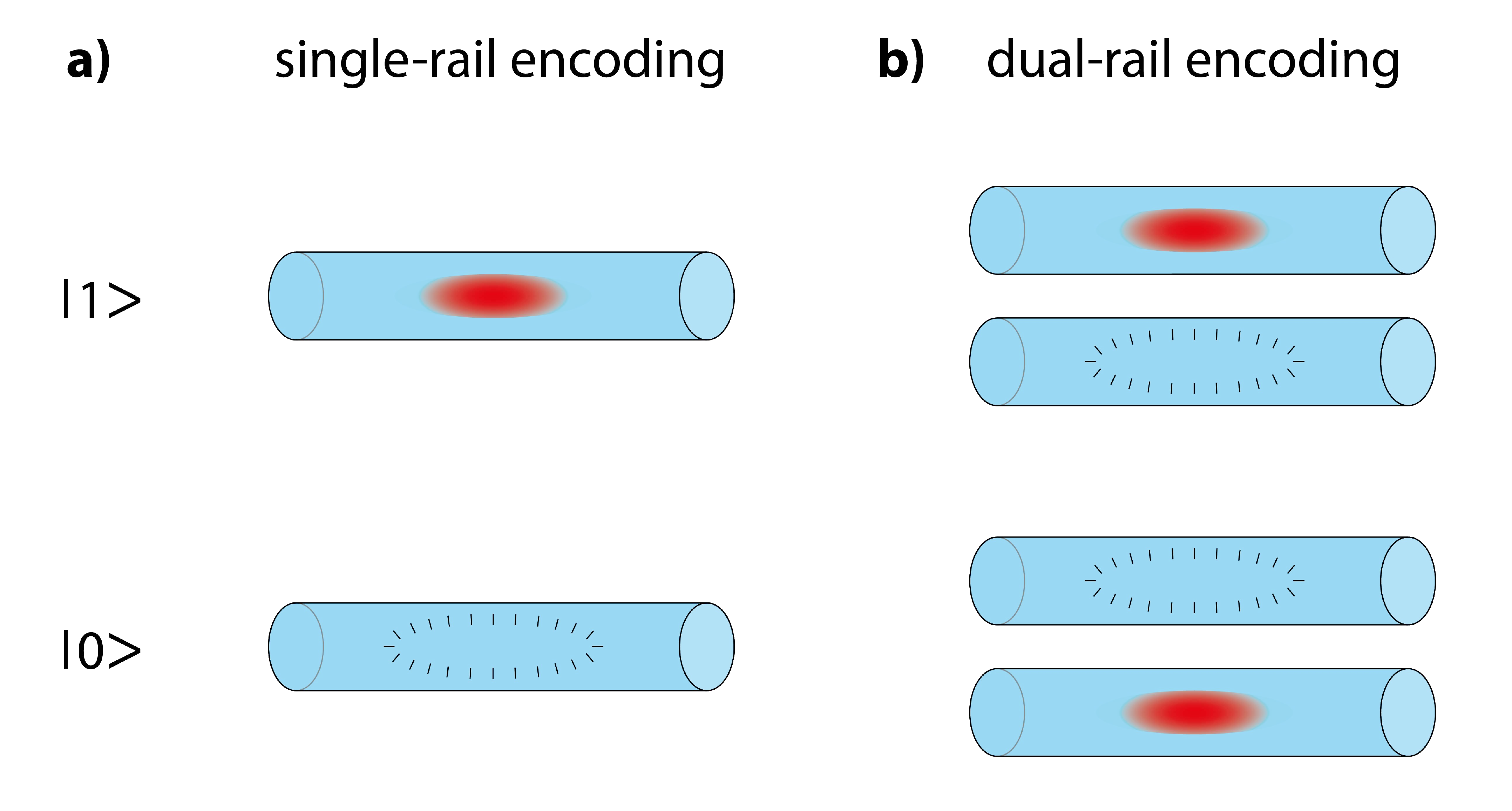}
\caption{{Encoding of optical qubits.}{(a)} Single-rail encoded qubits are represented by the absence or presence of a single photon of fixed polarization in one optical mode.
This encoding 
allows for the deterministic generation of entangled states,
 but with the caveat that single-qubit gates can only be achieved probabilistically. {(b)} Dual-rail encoded qubits are represented by the presence of a single photon in one or the other of two spatial optical modes.
Formally, polarization-encoded qubits are equivalent to the dual-rail encoding due to the basis comprising two orthogonal polarization modes, e.g., by defining $|1\rangle|0\rangle=|H\rangle$ and $|0\rangle|1\rangle=|V\rangle$ where $|H\rangle$ and $|V\rangle$ corresponds to a horizontally and vertically polarized photon state, respectively.}
\label{Fig1}
\end{figure*}

The subspace of interest in the single-rail encoding is spanned by the $6$-dimensional Fock state basis $\mathcal{A}\cup \mathcal{B}$ where
\bes
\label{eq:AB}
\begin{eqnarray}
\mc{A} &\equiv &\{1 ,a_{1}^{\dagger } ,a_{2}^{\dagger } ,a_{1}^{\dagger
}a_{2}^{\dagger } \}\vac, \label{A} \\
\mc{B} &\equiv &\{a_{1}^{\dagger 2},a_{2}^{\dagger
2}\} \vac/\sqrt{2} .
\label{B}
\end{eqnarray}
\ees
Here $\vac$ is the global vacuum,
$a_{i}^{\dagger }$, $a_{i}$, and $n_{i}$ are the creation,
annihilation, and number operator for mode $i$, respectively, satisfying the standard
bosonic commutation relations.
The subspace $\mc{A}$ is spanned by the computational basis of single-photon states in each of the two modes, while $\mathcal{B}$ is the subspace of {``bunched"}
two photon states in either one of the two modes (double-occupation states). The latter are the problematic states causing computational errors in the single-rail encoding.

To define a basis for the polarization (i.e., dual rail) encoding consider the creation operators $a_{ip}^\dagger$, where $i\in\{1,2\}$ is the spatial mode index, $p\in\{H,V\}$ is the polarization mode index, and all $a_{ip}^\dagger$ commute for all values of $i$ and $p$. The $i$th qubit basis states are then $\ket{0}_i=a_{iH}^\dagger\vac$ and $\ket{1}_i=a_{iV}^\dagger\vac$, so that the computational and double-occupation (error) subspaces are, respectively,
\bes
\begin{eqnarray}
\mc{A}' &\equiv &\{a_{1H}^{\dagger }a_{2H}^{\dagger } ,a_{1H}^{\dagger }a_{2V}^{\dagger }, a_{1V}^{\dagger }a_{2H}^{\dagger },a_{1V}^{\dagger }a_{2V}^{\dagger }\}\vac,
\label{A'} \\
\mc{B}' &\equiv &\{a_{iH}^{\dagger 2}/\sqrt{2},a_{iV}^{\dagger 2}/\sqrt{2},a_{iH}^{\dagger}a_{iV}^{\dagger}\}\vac,\quad i\in\{1,2\}
\label{B'}
\end{eqnarray}
\ees

It is well known that linear optics is insufficient for generating non-trivial (entangling) two-qubit gates in the dual-rail encoding. Indeed, this, together with the inherent weakness of non-linear photonic interactions such as the Kerr effect, is the {\it raison d'\^{e}tre} of the KLM scheme. The reason is a straightforward group-theoretic counting argument. Consider $m=2K$ optical modes. If dual-rail linear optics could generate an entangling gate between one pair of qubits, which could then be swapped to all other pairs of qubits using beam splitters, the set of
linear optical elements would generate the group U($2^{K}$). However, this contradicts the fact that the set of linear optical elements can only generate U($2K$)\cite {Reck:94,WuLidar}.
This means that it is impossible to generate an entangling gate in the subspace $\mc{A}'$, irrespective of what happens in $\mc{B}'$. {Another perspective on this is offered by results showing the unfeasibility of a linear optics Bell state analyzer, which rule out the implementation of a deterministic CNOT gate using dual-rail qubits\cite{Lutkenhaus:99a,Calsamiglia:01}.} However, these arguments cannot be applied in the single-rail case where a beam splitter suffices to generate two-qubit entanglement in the subspace $\mc{A}$. Since we are concerned in this work with deterministic gates we shall not consider the possibility of photon-bunching suppression and concurrent enhancement of the success probability of two-qubit gates in the dual-rail case. Instead,
we devote the remainder of this work to the no-go theorem for {deterministic gates in} the single-rail case.\\

\subsection{No-go theorem for two modes}
We now proceed to prove the following no-go theorem: it is not possible to implement a deterministic entangling gate between two single-rail qubits while at the same time removing the ``bunched" two-photon states using only linear optics. In this section we assume that only two photons in two modes are used; in the following section we allow for an arbitrary number of photons and modes.

The Hermitian beam-splitter generator is
\begin{equation}
X\equiv \frac{1}{2}\left( a_{2}^{\dagger }a_{1}+a_{1}^{\dagger }a_{2}\right) ,
\label{X}
\end{equation}%
and the Hermitian phase shifter generators are, for each mode,
\begin{equation}
n_{1}=a_{1}^{\dagger }a_{1},\quad n_{2}=a_{2}^{\dagger }a_{2}.
\end{equation}%
Let us also define
\begin{equation}
Y\equiv \frac{i}{2}\left( a_{2}^{\dagger }a_{1}-a_{1}^{\dagger }a_{2}\right)
,\qquad Z\equiv \frac{1}{2}(n_{1}-n_{2}).
\label{YZ}
\end{equation}%
Then it is easily checked
using the identities in the Methods section that the set of Hermitian operators $\{X,Y,Z\}$ is
closed under commutation and moreover satisfies su$(2)$ commutation
relations. Therefore the beam-splitter generator $X$ and the phase shifter
generators (combined as $Z$) generate U$(2)$. The most general unitary
evolution operator we can then construct according to the Euler decomposition is
the following composite gate:
\begin{equation}
U(\alpha ,\beta ,\gamma ,\delta ,\epsilon )=e^{i(\alpha n_{1}+\beta
n_{2})}e^{i\epsilon X}e^{i(\gamma n_{1}+\delta n_{2})},  \label{eq:U}
\end{equation}%
where we have allowed a more general linear combination of the two phase
shifter generators than $Z$, since they can be independently tuned.

We are interested
in the (reducible) representation of the unitary operator \eqref{eq:U} in
the $6$-dimensional {Fock state} basis $\mathcal{A}\cup \mathcal{B}$. We shall now show that if one demands that $U$ does not couple $\mc{A}$ and $\mc{B}$, then without further assumptions it is not possible to implement an entangling gate between the two {single-rail} qubits.}
In the $\mathcal{A}\cup \mathcal{B}$ basis we have
\begin{equation}
\begin{split}
& U(\alpha ,\beta ,\gamma ,\delta ,\epsilon )= \\
& \mathrm{diag}(1,e^{i\alpha },e^{i\beta },e^{i(\alpha +\beta )},e^{2i\alpha
},e^{2i\beta })\left[
\begin{array}{ccc}
1 &  &  \\
& A_{1} &  \\
&  & B_{1}%
\end{array}%
\right] \times  \\
& \mathrm{diag}(1,e^{i\gamma },e^{i\delta },e^{i(\gamma +\delta
)},e^{2i\gamma },e^{2i\delta }) \\
& \qquad =\left[
\begin{array}{ccc}
1 &  &  \\
& A_{2} &  \\
&  & B_{2}%
\end{array}%
\right] ,
\end{split}%
\end{equation}%
where diag denotes a diagonal matrix and where
\begin{equation}
\begin{split}
& A_{1}=\left[
\begin{array}{cc}
\cos \epsilon  & i\sin \epsilon  \\
i\sin \epsilon  & \cos \epsilon
\end{array}%
\right] , \\
& A_{2}=\left[
\begin{array}{cc}
e^{i(\alpha +\gamma )}\cos \epsilon  & ie^{i(\alpha +\delta )}\sin \epsilon
\\
ie^{i(\beta +\gamma )}\sin \epsilon  & e^{i(\beta +\delta )}\cos \epsilon
\end{array}%
\right] , \\
& B_{1}=\left[
\begin{array}{ccc}
\cos 2\epsilon  & \frac{1}{i\sqrt{2}}\sin 2\epsilon  & \frac{1}{i\sqrt{2}}%
\sin 2\epsilon  \\
\frac{1}{i\sqrt{2}}\sin 2\epsilon  & \cos ^{2}\epsilon  & -\sin ^{2}\epsilon
\\
\frac{1}{i\sqrt{2}}\sin 2\epsilon  & -\sin ^{2}\epsilon  & \cos ^{2}\epsilon
\end{array}%
\right] , \\
& B_{2}= \\
& \left[
\begin{array}{ccc}
e^{i(\alpha +\beta +\gamma +\delta )}\cos 2\epsilon  & e^{i(\alpha +\beta
+2\gamma )}\frac{\sin 2\epsilon }{i\sqrt{2}} & e^{i(\alpha +\beta +2\delta )}%
\frac{\sin 2\epsilon }{i\sqrt{2}} \\
e^{i(2\alpha +\gamma +\delta )}\frac{\sin 2\epsilon }{i\sqrt{2}} &
e^{2i(\alpha +\gamma )}\cos ^{2}\epsilon  & -e^{2i(\alpha +\delta )}\sin
^{2}\epsilon  \\
e^{i(2\beta +\gamma +\delta )}\frac{\sin 2\epsilon }{i\sqrt{2}} &
-e^{2i(\beta +\gamma )}\sin ^{2}\epsilon  & e^{2i(\beta +\delta )}\cos
^{2}\epsilon
\end{array}%
\right] .
\end{split}
\label{blocks}
\end{equation}%
Note that $U(\alpha ,\beta ,\gamma ,\delta ,\epsilon )$ consists of three
blocks ($1,A_{2}$, and $B_{2}$) of dimensions 1 (vacuum), 2 (one photon),
and 3 (two photons), respectively. Therefore repeated applications of the
composite gate $U$ with different values for the angles $\alpha ,\beta
,\gamma ,\delta ,\epsilon $ will still have the same triple block structure.
Hence our construction is
{general in the sense that} we need only consider a
single composite gate $U$.

Now, the problem is that the two-photon block includes $\mathcal{B}$. We
would like to decouple these two states from the other four. Clearly, this
can be done by setting the matrix elements $U_{4,5},U_{4,6},U_{5,4}$ and $%
U_{6,4}$
[i.e., the $(1,2),(1,3),(2,1),(3,1)$ elements of
$B_2$] to zero. As is clear from Eq.~\eqref{blocks}, a necessary and
sufficient condition for this is
\begin{equation}
\sin 2\epsilon =0.
\end{equation}%
This is achieved whenever $\epsilon $ is an integer multiple of $\pi /2$. If
we pick an even multiple $2n$ of $\pi /2$ we have%
\begin{eqnarray}
&U(\alpha ,\beta ,\gamma ,\delta ,n\pi )=\mathrm{diag}(1,(-1)^{n}e^{i(\alpha
+\gamma )},(-1)^{n}e^{i(\beta +\delta )},  \notag \\
&e^{i(\alpha +\beta +\gamma +\delta )},e^{2i(\alpha +\gamma )},e^{2i(\beta
+\delta )})  \label{eq:even}
\end{eqnarray}%
If we pick an odd multiple $2n+1$ of $\pi /2$ we have%
\begin{equation}
U(\alpha ,\beta ,\gamma ,\delta ,(2n+1)\frac{\pi }{2})=\left[
\begin{array}{ccc}
1 &  &  \\
& A_{3} &  \\
&  & B_{3}%
\end{array}%
\right] ,  \label{eq:odd}
\end{equation}%
where
\bes
\begin{eqnarray}
A_{3} &=&\left[
\begin{array}{cc}
0 & (-1)^{n}ie^{i(\alpha +\delta )} \\
(-1)^{n}ie^{i(\beta +\gamma )} & 0%
\end{array}%
\right] \\
B_{3} &=&\left[
\begin{array}{ccc}
-e^{i(\alpha +\beta +\gamma +\delta )} & 0 & 0 \\
0 & 0 & -e^{2i(\alpha +\delta )} \\
0 & -e^{2i(\beta +\gamma )} & 0%
\end{array}%
\right]
\end{eqnarray}
\ees
In both cases $U$ has a block-diagonal structure where the two-photons per
mode states are decoupled from the other four states. Considering just the
4-dimensional block acting on $\mathrm{span}(\mathcal{A}$), we have for the
even case%
\begin{eqnarray}
&\mathrm{diag}(1,(-1)^{n}e^{i(\alpha +\gamma )},(-1)^{n}e^{i(\beta +\delta
)},e^{i(\alpha +\beta +\gamma +\delta )})=  \notag \\
&\mathrm{diag}(1,(-1)^{n}e^{i(\beta +\delta )})\otimes \mathrm{diag}%
(1,(-1)^{n}e^{i(\alpha +\gamma )})
\end{eqnarray}%
i.e., a tensor product of two single-qubit phase gates. Similarly, for the
odd case we have

\begin{equation}
\begin{split}
&\left[
\begin{array}{cccc}
1 & 0 & 0 & 0 \\
0 & 0 & (-1)^{n}ie^{i(\alpha +\delta )} & 0 \\
0 & (-1)^{n}ie^{i(\beta +\gamma )} & 0 & 0 \\
0 & 0 & 0 & -e^{i(\alpha +\beta +\gamma +\delta )}%
\end{array}%
\right] \\
&=\textsc{SWAP}\, \mathrm{diag}(1,(-1)^{n}ie^{i(\alpha +\delta )})\otimes
\mathrm{diag}(1,(-1)^{n}ie^{i(\beta +\gamma )}),
\end{split}
\end{equation}
where $\textsc{SWAP} = \left[
\begin{smallmatrix}
1 & 0 & 0 & 0 \\
0 & 0 & 1 & 0 \\
0 & 1 & 0 & 0 \\
0 & 0 & 0 & 1%
\end{smallmatrix}
\right]$, a non-entangling gate. Unfortunately this means that, having decoupled $\mathrm{span}(\mathcal{B}$),
the action of $U$ on the $\mc{A}$ subspace is equivalent to a tensor product of two single-qubit phase gates, and therefore is no longer entangling.

The reason for this no-go result is straightforward: we are
forced to set $\epsilon $ equal to a multiple of $\pi /2$ in order to
prevent coupling to the $\mathrm{span}(\mathcal{B})$ subspace. Once we do
this we are left with only four independent parameters ($\alpha ,\beta
,\gamma ,\delta $), which are the parameters of the two phase shifters [see
Eq.~\eqref{eq:U}]. The latter can only generate phase gates.
To circumvent this result would require the use of some type of nonlinear
optical element (see Figure 2).

\begin{figure*}
\includegraphics[width=0.75\textwidth]{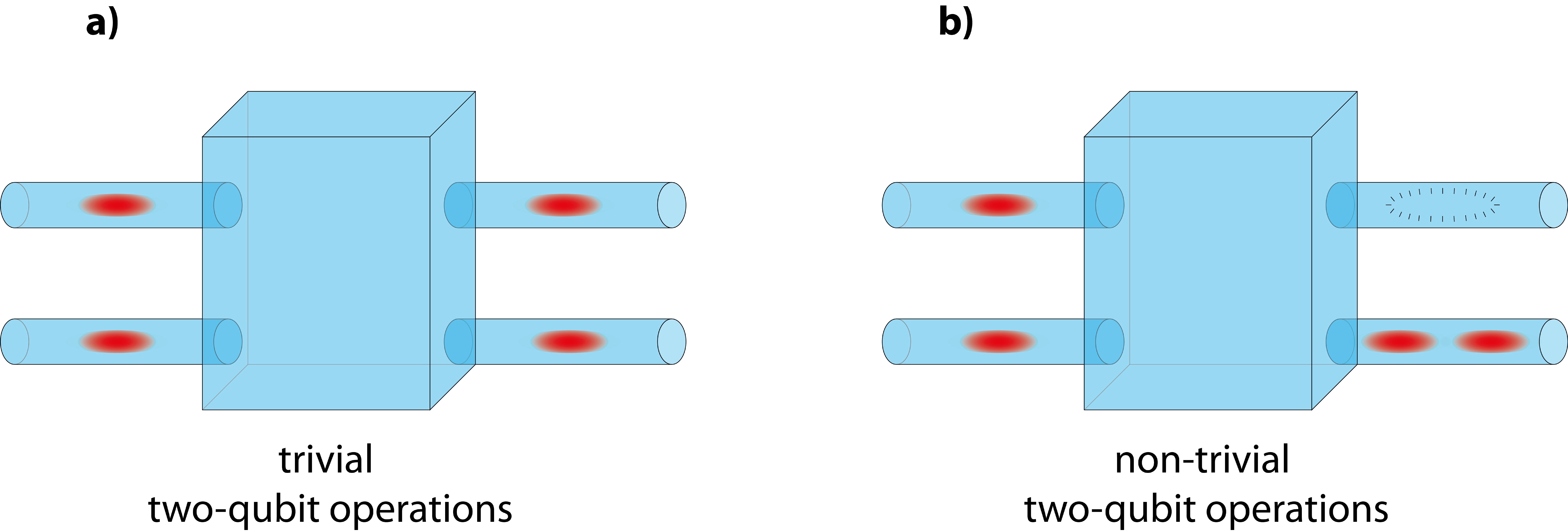}
\caption{{Two-qubit operations for single-rail encoded qubits.} (a) Only the restricted set of trivial two-qubit gates prohibits photon-bunching. {(b)} When using non-trivial two-qubit gates, such as the control-NOT or control-Phase gate, photon-bunching cannot be avoided and thus leads to gate errors or probabilistic success rates.}
\label{Fig2}
\end{figure*}

\subsection{No-go theorem for $M>2$ modes}
One might wonder
whether the inclusion of additional modes and photons allows to circumvent the no-go
theorem. In this section we show that this is not the case.

Consider a linear optical
system with $M$ modes containing
an arbitrary number of photons, of which only
two modes--supporting two qubits--are the target of linear
optical quantum computation.
We therefore refer to these two modes as
the ``computational modes.'' Clearly the choice of which two modes are identified as the computational modes is immaterial, so that without loss of generality we use the first two modes to this end.
The computational basis still comprises only those states having either zero or one photons per mode in the first two modes, except that now an arbitrary number of photons is allowed in the ancillary modes $3,\dots,M$. Let $\mc{C}_M \equiv \left\{ \prod_{j=1}^{M}\frac{1}{\sqrt{n_j!}}(a_{j}^{\dagger })^{n_{j}}\vac \right\}$, where the occupation numbers $\{n_j\}_{j=1}^M$ are arbitrary. Then the computational basis states are the set $\mc{A}_M = \left\{ \mc{C}_M\; | \; n_1,n_2\in\{0,1\}\right\}$. The basis set for the problematic subspace of bunched states is now $\mc{B}_M = \left\{ \mc{C}_M\; | \;  n_1,n_2\geq 2\right\}$. We assume that the photonic system is initialized in $\mathrm{span}(\mc{A}_M)$. Furthermore, we assume that the initial state is a product state between the computational and ancilla modes, i.e., there is no initial entanglement between these two set of modes. Our proof does not apply otherwise, so that initialization errors are outside of the scope of this work and will have to be dealt with by other means.

Now let $V$ be some arbitrary passive linear unitary optical transformation on all,
or some subset, of the $M$ modes. For example, $V$ could be a product of
phase shifters and beam splitters coupling any pair of modes.  Its most general form is
\begin{equation}
V=\exp \left(i\sum_{i,j=1}^{M}\theta _{ij}a_{i}^{\dagger }a_{j}\right) ,\quad \theta_{ij} = \theta_{ji} \in \mathbb{R},
\label{eq:V}
\end{equation}%
an arbitrary element of $U(M)$\cite{Reck:94}. The operator $V$ replaces $U$ [Eq.~\eqref{eq:U}] from our earlier two-mode discussion, and we shall show that even it cannot be used to decouple the computational basis states $\mc{A}_M$ from the bunched states $\mc{B}_M$ while simultaneously implementing an arbitrary unitary transformation on the computational modes. $V$ has the following unitary representation in terms of the $M\times M$ matrix $v = \{v_{ij}\}$, in the basis of bosonic creation operators (see Section~\ref{method}):
\begin{equation}
Va_{i}^{\dagger }V^{\dagger }=\sum_{j=1}^{M}v_{ij}a_{j}^{\dagger },\quad
i\in \{1,\dots ,M\} ,
\label{V}
\end{equation}
Note that $V$ commutes with the number operator, i.e., its only non-vanishing
matrix elements are between Fock states with equal occupation numbers. Note also that if we write $V = \exp(i \Omega)$ where $\Omega = \sum_{i,j=1}^{M}\theta _{ij}a_{i}^{\dagger }a_{j}$, then $\Omega\vac = 0$ due to the annihilation operator $a_j$, so that
{$V\vac =\vac$}, and hence $\langle {\bf 0}|V\vac =1$.

At first sight it might appear that the inclusion of ancilla modes could be useful. For example, one might hope that the bunched state $\ket{200}$ (for $M=3$) can be transformed into the computational state $\ket{002}$, where the two photons have been transferred to the ancilla mode. However, this approach fails for passive all-unitary linear optics. The reason is that we must rule out the reverse process, i.e., we must impose the following  \emph{necessary ``don't cause errors" condition}:
\begin{quote}
$V$ cannot transfer any photons from the ancillary modes into the computational modes in a manner which transforms a valid computational basis state into a bunched state.
\end{quote}
For example, $V$ is not allowed to take the computational basis state $(a_{3}^{\dagger })^2\vac \in \mc{A}_M$ to $(a_{1}^{\dagger })^2\vac \in \mc{B}_M$ (ignoring normalization). Thus, we impose the necessary conditions for all $i\in\{3,\dots,M\}$
\bes
\begin{align}
& \vacb a_{1}^{2}V(a_{i}^{\dagger })^2\vac  =0 , \\
& \vacb a_{2}^{2}V(a_{i}^{\dagger })^2\vac  =0
\end{align}
\ees
Then,
\bes
\begin{align}
0 &=\vacb a_{1}^{2}V(a_{i}^{\dagger })^2\vac  =\vacb
a_{1}^{2}Va_{i}^{\dagger }V^{\dagger }Va_{i}^{\dagger }V^{\dagger}V\vac    \\
&=\sum_{k,l=1}^{M}v_{ik}v_{il}\vacb
a_{1}^{2}a_{k}^{\dagger }a_{l}^{\dagger }V\vac \label{18b} \\
&=v_{i1}^2\vacb
a_{1}^{2}a_{1}^{\dagger }a_{1}^{\dagger }\vac +\sum_{k,l=2}^{M}v_{ik}v_{il}\vacb
a_{k}^{\dagger }a_{l}^{\dagger }a_{1}^{2}\vac \notag
\\
&\qquad+ 2\sum_{k=2}^{M}v_{ik}v_{i1}\vacb
a_{k}^{\dagger }a_{1}^{2}a_{1}^{\dagger }\vac
\label{18c} \\
& = v_{i1}^2\vacb
a_{1}^{2}a_{1}^{\dagger }a_{1}^{\dagger }\vac
\label{18d}
\end{align}
\ees
where to arrive at \eqref{18b} we used Eq.~\eqref{V}, and to arrive at \eqref{18c} we used $V\vac=\vac$ along with the standard bosonic commutation relations, 
and to arrive at \eqref{18d} we used the fact that both sums in \eqref{18c} vanish due to $\vacb a_k^\dagger = 0$. We can thus conclude that $v_{i1}$ must vanish for all $i\in\{3,\dots,M\}$. After we impose the additional necessary condition $\vacb a_{2}^{2}V(a_{i}^{\dagger })^2\vac  =0$, we find that $v_{i2}$ must vanish for all $i\in\{3,\dots,M\}$. In addition, it follows from unitarity of $v$ that then also  $v_{1i}$ and $v_{2i}$ must vanish for all $i\in\{3,\dots,M\}$ (see Section~\ref{method}). That is, we have proved that the ``don't cause errors" condition implies that $v$ has a block-diagonal structure:
\beq
v = v_c \oplus v_a ,
\eeq
where $v_c$ is a $2\times 2$ block (over the computational modes $i,j \in \{1,2\}$) and $v_a$ is an $(M-2)\times (M-2)$ block (over the ancilla modes $i,j \in \{3,\dots,M\}$). This, in turn, can be interpreted as ``don't couple". It states that, subject to the ``don't cause errors" condition, $V$ cannot couple the computational and ancilla modes, and in particular cannot cause photons to leak from the ancilla modes into the computational, or \textit{vice versa}. For example, $V$ cannot couple the two computational basis states $\ket{001}$ and $\ket{100}$ ($M=3$). It is conceptually clear that this implies that nothing is gained by the introduction of the ancilla modes. However, let us provide a formal proof as well.

Assume, as before, that the system is
initially in a factorized state between the computational and ancilla modes: $G(a_{3}^{\dagger },\dots,a_{M}^{\dagger})F(a_{1}^{\dagger },a_{2}^{\dagger })\vac$, where
$F$ and $G$ are arbitrary polynomials in the creation operators, e.g., $F(a_{1}^{\dagger },a_{2}^{\dagger }) = \sum_{m_1,m_2=0}c_{m_1,m_2} (a_1^\dagger)^{m_1}(a_2^\dagger)^{m_2}$, where $c_{m_1,m_2}$ are arbitrary coefficients.  Note that because of the block-diagonal structure of $v$ it follows that
\begin{align}
&VF(a_{1}^{\dagger },a_{2}^{\dagger })V^\dagger = \sum_{m_1,m_2=0}c_{m_1,m_2} (Va_1^\dagger V^\dagger)^{m_1}(Va_2^\dagger V^\dagger)^{m_2} \notag \\
&\quad  = \sum_{m_1,m_2=0}c_{m_1,m_2} \left(\sum_{j=1}^2 v_{1j}a_j^\dagger\right)^{m_1}\left(\sum_{j'=1}^2 v_{2j'}a_{j'}^\dagger\right)^{m_2} \notag \\
&\quad = F'(a_{1}^{\dagger },a_{2}^{\dagger }) ,
\end{align}
where $F'$ is a new polynomial. Here the key point is that the ancilla modes do not appear in $F'$. A similar calculation reveals that $V G(a_{3}^{\dagger },\dots,a_{M}^{\dagger }) = G(Va_{3}^{\dagger }V^\dagger,\dots,Va_{M}^{\dagger }V^\dagger)= G'(a_{3}^{\dagger },\dots,a_{M}^{\dagger})$, where $G'$ is a new polynomial. Therefore the action of $V$ on an arbitrary factorized initial state is
\bes
\begin{align}
V& G(a_{3}^{\dagger },\dots,a_{M}^{\dagger })F(a_{1}^{\dagger
},a_{2}^{\dagger })\vac \\
&=[V G(a_{3}^{\dagger },\dots,a_{M}^{\dagger })V^{\dagger }][VF(a_{1}^{\dagger},a_{2}^{\dagger })V^{\dagger }]V\vac \\
& =G'(a_{3}^{\dagger },\dots,a_{M}^{\dagger})F'(a_{1}^{\dagger },a_{2}^{\dagger })\vac .
\end{align}
\ees
For the computational
modes this is equivalent to $VF(a_{1}^{\dagger },a_{2}^{\dagger })\vac=F'(a_{1}^{\dagger },a_{2}^{\dagger })\vac$, where the no-go theorem for $M=2$ holds.\\

\section{Discussion}
We have shown that using linear optics alone it is not possible to
cancel the photon bunching effect while at the same time implementing a deterministic universal set of logic gates using single-rail photonic qubits. In spirit our result agrees with previous theoretical
work concerning the dual-rail encoding case\cite{Lutkenhaus:99a,Knill:00,Calsamiglia:01}, showing that passive linear optics does not involve particle interactions other than those imposed by statistics, and can be understood in terms of classical wave mechanics. One approach is to then consider additional nonlinear operations such as photon-detection or absorption\cite{Franson:04}, the Kerr effect, or light squeezing,
in order to enable universal  linear optical quantum computing.
However, our work leaves open the possibility that
a more general analysis than we have considered here, in particular one which accounts for the possibility of a \emph{probabilistic} enhancement (or even optimization) of both photon-bunching suppression and linear optical gate fidelity, might circumvent our no-go theorem. Nor did we consider measurements of the ancilla modes or other non-unitary transformations. For example, one might consider a non-unitary transformation with maps the bunched state $\ket{200}$ (on a total of three modes) to the computational basis state $\ket{002}$, without at the same time mapping $\ket{002}$ back into the computational subspace. In this sense the results presented here are thus a starting point for a study of the use of additional resources in linear optics quantum computing using single-rail qubits.

\section{Methods}
\label{methods}
\subsection{Unitary representation of $V$ in the creation operator basis}
To prove that the representation of $V$ in Eq.~\eqref{V}  is unitary one can use the standard  bosonic commutation relations
along with the  Baker-Hausdorff formula
\beq
e^{\alpha A}Be^{-\alpha A}=B+\sum_{m=1}^\infty \frac{\alpha^{m}}{m!}[_mA,B] ,
\eeq
where $[_m A,B]:=[A,[_{m-1}A,B]]$ and $[_1A,B]\equiv[A,B]$, for arbitrary operators $A$ and $B$, and $\alpha\in\mathbb{C}$, to establish that
$v = e^{-i\theta}$, where $\theta = \{\theta_{ij}\}$ is the orthogonal matrix of angles appearing in Eq.~\eqref{eq:V}.  Indeed, setting $\alpha =i$, $A=\sum_{kj}\theta _{kj}a_{k}^{\dag }a_{j}$, $%
B=a_{i}^{\dag }$, and using the identity $[a_k^\dag a_j,a_i^\dag] = \delta_{ij} a_k^\dag$,
we have:
\bes
\begin{align}
\lbrack A,B] &=\sum_{j}[\theta ]_{ji}a_{j}^{\dag } \\
\lbrack _{2}A,B] &=\sum_{jkm}\theta _{jk}\theta _{mi}[a_{j}^{\dag
}a_{k},a_{m}^{\dag }]=\sum_{jm}\theta _{jm}\theta _{mi}a_{j}^{\dag
}\notag \\
&=\sum_{j}[\theta ^{2}]_{ji}a_{j}^{\dag },
\end{align}
\ees
etc., so that $[_{m}A,B]=\sum_{j=1}^M [\theta ^{m}]_{ji}a_{j}^{\dag }$. Thus
\begin{align}
Va_{i}^{\dag }V^{\dag } &= \sum_{m=0}^{\infty }\frac{i^{m}}{m!} \sum_{j=1}^M[\theta ^{m}]_{ji}a_{j}^{\dag }
= \sum_{j=1}^M\left[ e^{i\theta }\right] _{ji}a_{k}^{\dag },
\end{align}
so that $v_{ij} = \left[ e^{i\theta }\right] _{ji}$, whence $v=e^{-i\theta }$ as claimed.

\subsection{Proof that ``don't cause errors" implies ``don't couple"}
Consider a unitary matrix $v$ with the block structure
\beq
\label{eq:v-block}
v=\left(
\begin{array}{ll}
x & y \\
0 & z%
\end{array}%
\right) ,
\eeq
where $x$ and $z$ are square and $y$ can be rectangular. The unitarity condition $v^{\dag }v=I$ yields
\beq
\left(
\begin{array}{ll}
x^{\dag } & 0 \\
y^{\dag } & z^{\dag }%
\end{array}%
\right) \left(
\begin{array}{ll}
x & y \\
0 & z%
\end{array}%
\right) =\left(
\begin{array}{ll}
x^{\dag }x & x^{\dag }y \\
y^{\dag }x & y^{\dag }y+z^{\dag }z%
\end{array}%
\right) =\left(
\begin{array}{ll}
I & 0 \\
0 & I%
\end{array}%
\right) ,
\eeq
so that $x^{\dag }x=I$ and $x^{\dag }y=0$. By unitarity of $v$ the matrix $x$ cannot be zero, so $%
x^{-1}=x^{\dag }$ and hence also $xx^{\dag }=I$. Thus $xx^{\dag }y=y=0$. In the context of our proof in the last section, we showed that the ``don't cause errors" condition implies that both $v_{i1}=v_{i2}=0$ for all $i\in\{3,\dots,M\}$. This is the $0$ block in Eq.~\eqref{eq:v-block}. The vanishing of the $y$ block then implies that also $v_{1i}=v_{2i}=0$ for all $i\in\{3,\dots,M\}$, which is the ``don't couple" result.

\acknowledgments
L.-A.W. is supported by the Ikerbasque Foundation Start-up,
the Basque Government (grant IT472-10) and the Spanish MEC (Project No.
FIS2009-12773-C02-02). P.W. acknowledges support from the European Commission, Q-ESSENCE (No. 248095), QUILMI (No. 295293) and the ERA-Net CHISTERA project QUASAR, the John Templeton Foundation, the Vienna Center for Quantum Science and Technology (VCQ), the Austrian Nano-initiative NAP Platon, the Austrian Science Fund (FWF) through the SFB FoQuS (No. F4006-N16), START (No. Y585-N20) and the doctoral programme CoQuS, the Vienna Science and Technology Fund (WWTF) under grant ICT12-041, and the Air Force Office of Scientific Research, Air Force Material Command, United States Air Force, under grant number FA8655-11-1-3004. D.A.L. is supported by the National Science Foundation under grant No. PHY-969969 and by the ARO MURI
grant W911NF-11-1-0268.

\end{document}